\documentclass[aps, prl, twocolumn, superscriptaddress, amsmath, showpacs, tightenlines, footinbib, longbibliography]{revtex4-1}
\usepackage[colorlinks, breaklinks, citecolor=blue, linkcolor=blue,  urlcolor={blue}]{hyperref}
\usepackage{breakurl}
\usepackage{amsfonts}
\usepackage{amsmath}
\usepackage{amssymb}
\usepackage{mathrsfs}
\usepackage{epsfig}
\usepackage{graphicx}
\usepackage{bm}

\begin{document}

\title{Nonreciprocal transition between two nondegenerate energy levels}
\author{Xun-Wei Xu}
\email{davidxu0816@163.com}
\affiliation{Department of Applied Physics, East China Jiaotong University, Nanchang,
330013, China}
\author{Yan-Jun Zhao}
\affiliation{Faculty of Information Technology, College of Microelectronics, Beijing
University of Technology, Beijing, 100124, China}
\author{Hui Wang}
\affiliation{Center for Emergent Matter Science (CEMS), RIKEN, Wako, Saitama 351-0198, Japan}
\author{Ai-Xi Chen}
\email{aixichen@zstu.edu.cn}
\affiliation{Department of Physics, Zhejiang Sci-Tech University, Hangzhou, 310018, China}
\affiliation{Department of Applied Physics, East China Jiaotong University, Nanchang,
330013, China}
\author{Yu-xi Liu}
\email{yuxiliu@tsinghua.edu.cn}
\affiliation{Institute of Microelectronics, Tsinghua University, Beijing 100084, China}
\affiliation{Beijing National Research Center for Information Science and Technology
(BNRist), Beijing 100084, China}
\date{\today }

\begin{abstract}
Stimulated emission and absorption are two fundamental processes of light-matter interaction, and the coefficients of the two  processes should be equal in general.
However, we will describe a generic method to realize significant difference between the
stimulated emission and absorption coefficients of two nondegenerate energy levels,
which we refer to as nonreciprocal transition.
As a simple implementation, a cyclic three-level atom system, comprising two nondegenerate energy levels and one auxiliary energy level, is employed to show nonreciprocal transition via a combination of synthetic magnetism and reservoir engineering.
Moreover, a single-photon nonreciprocal transporter is proposed using two one dimensional semi-infinite coupled-resonator waveguides connected by an atom with nonreciprocal transition effect.
Our work opens up a route to design atom-mediated nonreciprocal devices
in a wide range of physical systems.
\end{abstract}

\maketitle



\emph{Introduction.---} According to Einstein's phenomenological radiation
theory~\cite{Einstein1917}, the absorption coefficient should be equal to
the stimulated emission coefficient between two nondegenerate energy levels. When
the spontaneous emission can be neglected, a two-level system undergoes
optical Rabi oscillations under the action of a coherent driving
electromagnetic field~\cite{ScullyBook}. However, can we make the absorption
coefficient different from the stimulated emission coefficient for the transition between
two energy levels with different eigenvalues, i.e., nonreciprocal transition
between two nondegenerate energy levels? The answer is YES. In this paper, we
describe a generic method to realize nonreciprocal transition between two
nondegenerate energy levels, and show that the absorption and stimulated emission coefficients can be controlled via a combination of synthetic magnetism and reservoir engineering.

A theoretical research showed that~\cite{MetelmannPRX15} a combination of synthetic magnetism and reservoir engineering can be used to implement nonreciprocal photon transmission and amplification in coupled photonic systems, and this has been confirmed by a recent experiment~\cite{KFangNPy17}.
Based on a similar mechanism, many different schemes for nonreciprocal photon
transport are proposed theoretically~\cite{XuXWPRA15,XWXuPRA16a,MetelmannarX16a,LTianPRA17} and implemented experimentally~\cite{RuesinkNC16a,PetersonPRX17,BernierNC17,BarzanjehNC17}.
Synthetic magnetism is an effective approach to achieve non-reciprocal transport of
uncharged particles, such as photons~\cite%
{KochPRA10,UmucalilarPRA11,YPWangSR15,FXSunNJP17} or phonons~\cite%
{HabrakenNJP12,SeifNC18}, for the potential applications in simulating
quantummany-body phenomena~\cite%
{RechtsmanNat13,SchmidtOpt15,PeanoPRX15,PeanoPRX16,PeanoNC16,MinkovOptica16,BrendelPRB18}%
, and creating devices robust against disorder and backscattering~\cite%
{HafeziNPy11,KFangNPo12,TzuangNPo14,SliwaPRX15}.
Reservoir engineering~\cite{PoyatosPRL96} has been a significant subject for generating useful quantum
behavior by specially designing the couplings between a system of interest
and a structured dissipative environment, such as cooling mechanical
harmonic oscillators~\cite{XXuPRL17}, synthesise quantum harmonic oscillator
state~\cite{KienzlerSci15}, and generating state-dependent photon blockade~%
\cite{MiranowiczPRA14}, stable entanglement between two nanomechanical
resonators~\cite{CJYangPRA15,XBYanPRA17}, and squeezed states of
nanomechanical resonators~\cite{RablPRB04,WoolleyPRA14}.

In this paper, we introduce the concept of nonreciprocity to investigate the
transitions between different energy levels, and generalize the general strategy for nonreciprocal photon
transmission~\cite{MetelmannPRX15} to the atomic systems to achieve nonreciprocal transition between two nondegenerate energy levels.
As a simple implementation, a cyclic three-level atom system, comprising two nondegenerate energy levels and one auxiliary energy level, is employed to show nonreciprocal transition via a combination of synthetic magnetism and reservoir engineering.

In application, the atomic systems with nonreciprocal transition allow one to generate nonreciprocal devices.
In this paper, a single-photon nonreciprocal transporter is proposed in a system of two one dimensional semi-infinite coupled-resonator waveguides connected by an atom based on the nonreciprocal transition effect.
The nonreciprocal transition effect provides a new routine to design atom-mediated nonreciprocal devices in a verity of physical systems.

\begin{figure}[tbp]
\includegraphics[bb=41 149 549 590, width=7.5 cm, clip]{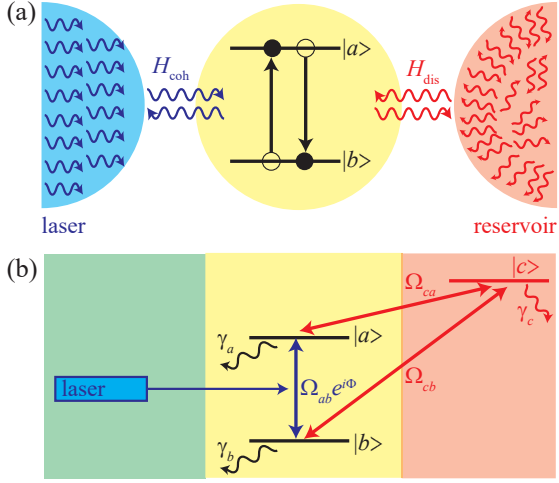}
\caption{(Color online) (a) Schematic diagram for generating nonreciprocal
transition: two nondegenerate energy levels $\left\vert a\right\rangle $ and $%
\left\vert b\right\rangle $ are coupled to one another via a coherent
interaction $H_{\mathrm{coh}}$, and are also coupled to the same engineered
reservoir. (b) Schematic diagram for implementation of nonreciprocal
transition in a cyclic three-level atom (characterized by $\left\vert a\right\rangle $, $%
\left\vert b\right\rangle $, and $\left\vert c\right\rangle $). A laser field ($\Omega _{ab}e^{i\Phi}$) is applied to drive the direct transition between the two levels $\left\vert a\right\rangle $ and $\left\vert b\right\rangle $, and they are also coupled indirectly by the auxiliary level $\left\vert c\right\rangle $ through two laser fields ($\Omega _{ca}$ and $\Omega _{cb}$),
where the decay of the level $\left\vert c\right\rangle $ is much faster than the other two levels,
i.e., $\gamma _{c}\gg \max \left\{ \gamma _{a},\gamma _{b}\right\} $, so
that the auxiliary level $\left\vert c\right\rangle $ is served as a engineered
reservoir.}
\label{fig1}
\end{figure}

\emph{General method for nonreciprocal transition.---} A general
model of two nondegenerate energy levels $\left\vert a\right\rangle $ and $%
\left\vert b\right\rangle $ for nonreciprocal transition is shown in Fig.~\ref{fig1}(a). There are two distinct ways through which the two levels are coupled
to one another. The first one is they are coupled through a coherent
interaction $H_{\mathrm{coh}}$, which is described by $H_{\mathrm{coh}%
} =\Omega \left\vert a\right\rangle \left\langle b\right\vert +\Omega
^{*}\left\vert b\right\rangle \left\langle a\right\vert$ with coupling strength $\Omega$. The simplest
implementation of the coherent interaction $H_{\mathrm{coh}}$ is driving the
two levels with a coherent laser field.

The second way is that they are coupled to the same engineered reservoir. A
dissipative interaction $H_{\mathrm{dis}}$ between the two levels can be
obtained by adiabatically eliminating the engineered reservoir. The
effective Hamiltonian for the dissipative interaction $H_{\mathrm{dis}}$ can
be written in a non-Hermitian form as $H_{\mathrm{dis}} = -i\gamma (\left\vert
a\right\rangle \left\langle b\right\vert +\left\vert
b\right\rangle \left\langle a\right\vert)$ with coupling strength $\gamma$. This dissipative version of
interaction can be implemented by an auxiliary energy level, which is
damping much faster than the two levels. The details on the realization will be
shown in the next section.

Base on the two distinct ways, the total Hamiltonian for the interaction
between the two levels are
\begin{equation}
H_{\mathrm{coh+dis}} = (\Omega-i\gamma) \left\vert a\right\rangle
\left\langle b\right\vert +(\Omega^{*}-i\gamma)\left\vert b\right\rangle
\left\langle a\right\vert.
\end{equation}%
When $\Omega=-i\gamma$ and $\Omega^{*}\neq \Omega$, there is only transition $\left\vert a\right\rangle
\rightarrow \left\vert b\right\rangle$ but $\left\vert b\right\rangle
\nrightarrow \left\vert a\right\rangle$. Instead, when $\Omega^{*}=i\gamma$ and $\Omega^{*}\neq \Omega$,
there is only transition $\left\vert b\right\rangle \rightarrow \left\vert
a\right\rangle$ but $\left\vert a\right\rangle \nrightarrow \left\vert
b\right\rangle$.

\emph{Nonreciprocal transition with cyclic three-level transition.---} To
make the method more concrete, we show how to implement nonreciprocal
transition in a cyclic three-level atom, as depicted in Fig.~\ref{fig1}(b). We
consider a cyclic three-level atom ($\left\vert a\right\rangle $, $%
\left\vert b\right\rangle $, and $\left\vert c\right\rangle $) driven by
three classical coherent fields (at rates $\Omega _{ij}$, frequencies $\nu
_{ij}$, phases $\phi _{ij}$, with $i,j=a,b,c$) that is described by a
Hamiltonian $H=H_{0}+H_{1}$ as~\cite{SM}
\begin{equation}
H_{0}=\left( \Delta _{ab}-i\gamma _{a}\right) \left\vert a\right\rangle
\left\langle a\right\vert -i\gamma _{b}\left\vert b\right\rangle
\left\langle b\right\vert +\left( \Delta _{cb}-i\gamma _{c}\right)
\left\vert c\right\rangle \left\langle c\right\vert ,
\end{equation}%
\begin{equation}
H_{1}=\Omega _{ab}e^{i\Phi }\left\vert a\right\rangle
\left\langle b\right\vert +\Omega _{cb}\left\vert c\right\rangle
\left\langle b\right\vert +\Omega _{ca}\left\vert c\right\rangle
\left\langle a\right\vert +\mathrm{H.c.},
\end{equation}%
where $\Delta _{ij}=\omega _{ij}-\nu _{ij}$ $\left( i,j=a,b,c\right) $, $%
\omega _{ij}$ is the frequency difference between levels $\left\vert
i\right\rangle $ and $\left\vert j\right\rangle $; $\gamma _{i}$ ($i=a,b,c$) is the decay rates. We assume that $\nu
_{ab}=\nu _{cb}-\nu _{ca}$, so the detuning $\Delta _{ab}=\Delta
_{cb}-\Delta _{ca}$. The synthetic magnetic flux $\Phi \equiv\phi _{ab}+\phi
_{bc}+\phi _{ca}$ is the total phase of the three driving fields around the
cyclic three-level atom and independent of the local redefinition of the
states $\left\vert i\right\rangle $. We note that the synthetic magnetic flux $\Phi $
inducing circulation of population between three energy levels has been observed in
a recent experiment~\cite{BarfussArx18}. Nevertheless, we consider the decays
of the three levels, and the decay of
level $\left\vert c\right\rangle $ is much faster than the other two levels,
i.e., $\gamma _{c}\gg \max \left\{ \gamma _{a},\gamma _{b}\right\} $, so
that level $\left\vert c\right\rangle $ is served as an engineered
reservoir.

In order to show the nonreciprocal transition between levels $\left\vert
a\right\rangle $ and $\left\vert b\right\rangle $ intuitively, we can derive
an effective Hamiltonian by eliminating the level $\left\vert c\right\rangle
$ (the engineered reservoir) adiabatically (see the Supplemental Material~%
\cite{SM}), under the assumption that $\gamma _{c}\gg \max \left\{ \gamma
_{a},\gamma _{b}\right\} $. Then an effective (non-Hermitian) Hamiltonian only including
levels $\left\vert a\right\rangle $ and $\left\vert b\right\rangle $ is
given by
\begin{eqnarray}  \label{eq4}
H_{\mathrm{eff}} &=&\left( \Delta _{a}-i\Gamma _{a}\right) \left\vert
a\right\rangle \left\langle a\right\vert +\left( \Delta _{b}-i\Gamma
_{b}\right) \left\vert b\right\rangle \left\langle b\right\vert  \nonumber \\
&&+J_{ab}\left\vert a\right\rangle \left\langle b\right\vert
+J_{ba}\left\vert b\right\rangle \left\langle a\right\vert ,
\end{eqnarray}
with detuning $\Delta _{a}\equiv \Delta _{ab}-\Omega _{ca}^{2}\Delta
_{cb}/(\gamma _{c}^{2}+\Delta _{cb}^{2})$ and $\Delta _{b}\equiv -\Omega
_{cb}^{2}\Delta _{cb}/(\gamma _{c}^{2}+\Delta _{cb}^{2})$, effective decay
rates $\Gamma _{a}\equiv \gamma _{a}+\Omega _{ca}^{2}\gamma _{c}/(\gamma
_{c}^{2}+\Delta _{cb}^{2})$ and $\Gamma _{b}\equiv \gamma _{b}+\Omega
_{cb}^{2}\gamma _{c}/(\gamma _{c}^{2}+\Delta _{cb}^{2})$, and effective
coupling coefficients $J_{ab}\equiv \Omega _{ab}e^{i\Phi }-i\Omega
_{ca}\Omega _{cb}\left( \gamma _{c}-i\Delta _{cb}\right) /(\gamma
_{c}^{2}+\Delta _{cb}^{2})$ and $J_{ba}\equiv \Omega _{ab}e^{-i\Phi
}-i\Omega _{ca}\Omega _{cb}\left( \gamma _{c}-i\Delta _{cb}\right) /(\gamma
_{c}^{2}+\Delta _{cb}^{2})$, which include the coherent interaction $\Omega
_{ab}e^{\pm i\Phi }$ coming from the coherent driving field, and the
dissipative interaction $-i\Omega _{ca}\Omega _{cb}\left( \gamma
_{c}-i\Delta _{cb}\right) /(\gamma _{c}^{2}+\Delta _{cb}^{2})$ induced by
the axillary level $\left\vert c\right\rangle $. The effective coupling
coefficients $J_{ab}$ and $J_{ba}$ are dependent on the synthetic magnetic
flux $\Phi $. Under the resonant condition $\Delta _{cb}=0$, we have $%
|J_{ab}|<|J_{ba}|$ for $0<\Phi <\pi $; conversely, we have $%
|J_{ab}|>|J_{ba}| $ as $-\pi <\Phi <0$. Thus, the perfect nonreciprocal
transition, i.e, $J_{ab}=0$ and $J_{ba}\neq 0$ (or $J_{ba}=0$ and $%
J_{ab}\neq 0$), is obtained when $\Phi =\pi /2$ (or $\Phi =-\pi /2$) with $%
\Omega _{ab}=\Omega _{ca}\Omega _{cb}/\gamma _{c}$.

\begin{figure}[tbp]
\includegraphics[bb=63 191 513 561, width=8.5 cm, clip]{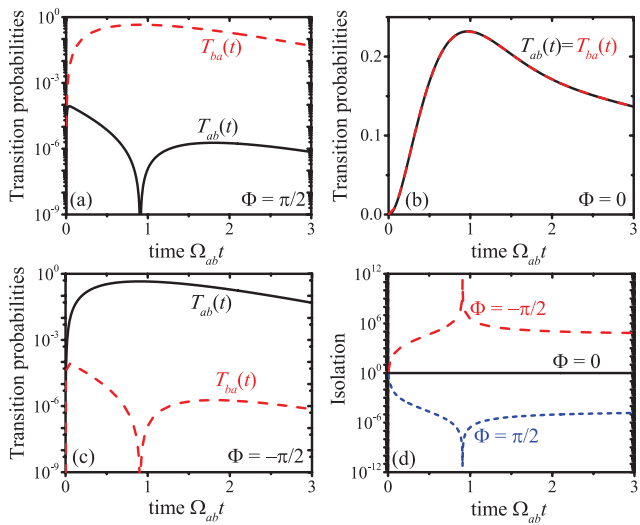}
\caption{(Color online) The transition probabilities $T_{ab}(t)$ and $%
T_{ba}(t)$ are plotted as functions of the time $\Omega _{ab} t$ for: (a) $\Phi=%
\protect\pi/2$; (b) $\Phi=0$; (c) $\Phi=-\protect\pi/2$. (d) The isolation $%
I(t)$ is plotted as a function of time $\Omega _{ab} t$ for $\Phi=(\protect\pi%
/2,0,-\protect\pi/2)$. The other parameters are $\protect\gamma _{a}=\protect%
\gamma _{b}=\Omega _{ab}/10$, $\protect\gamma _{c}=100\Omega _{ab}$, $\Omega
_{ca}=\Omega _{bc}=10\Omega _{ab}$, and $\Delta
_{cb}=\Delta_{ca}=\Delta _{ab}=0$.}
\label{fig2}
\end{figure}

To understand the nonreciprocal transition further, we  take a view on the dynamical behavior of the transition probabilities between levels $|a\rangle
$ and $|b\rangle $. The time-evolution operator for the Hamiltonian $H$ is
given by $U(t)=\exp (-iHt)$, and the probabilities for transitions $%
|a\rangle \rightarrow |b\rangle $ and $|b\rangle \rightarrow |a\rangle $ can
be defined by $T_{ba}(t)\equiv |\left\langle a\right\vert U(t)|b\rangle
|^{2} $ and $T_{ab}(t)\equiv |\left\langle b\right\vert U(t)|a\rangle |^{2}$%
, respectively. They are plotted as functions of time $\Omega _{ab} t$ in Figs.~%
\ref{fig2}(a)-\ref{fig2}(c). It is clear that $T_{ab}(t)\ll T_{ba}(t)$ for $%
\phi =\pi /2 $, $T_{ab}(t)\gg T_{ba}(t)$ for $\phi =-\pi /2$, and $%
T_{ab}(t)= T_{ba}(t)$ for $\phi =0$. The isolation for the nonreciprocal
transition defined by $I(t)\equiv T_{ab}(t)/T_{ba}(t)$ is plotted as a
function of time $\Omega _{ab} t$ in Fig.~\ref{fig2}(d). One can achieve $%
I(t)>10^{6}$ for $\Phi =-\pi /2$ and $I(t)<10^{-6}$ for $\Phi =\pi /2$ at
time $\Omega _{ab} t=1$.

\begin{figure}[tbp]
\includegraphics[bb=108 288 470 645, width=8.5 cm, clip]{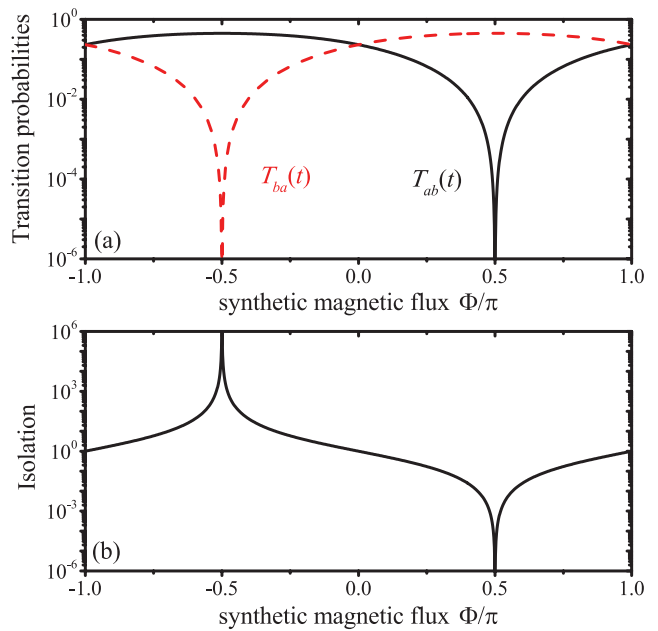}
\caption{(Color online) (a) The transition probabilities $T_{ab}(t)$ and $%
T_{ba}(t)$ and (b) The isolation $I(t)$ are plotted as functions of the
synthetic magnetic flux $\Phi$ at time $\Omega _{ab} t=1$. The other parameters
are $\protect\gamma _{a}=\protect\gamma _{b}=\Omega _{ab}/10$, $\protect\gamma %
_{c}=100\Omega _{ab}$, $\Omega _{ca}=\Omega _{bc}=10\Omega _{ab}$, and $\Delta _{cb}=\Delta_{ca}=\Delta _{ab}=0$.}
\label{fig3}
\end{figure}

Furthermore, the dependence of the transition probabilities $T_{ab}(t)$ and $%
T_{ba}(t)$ on the synthetic magnetic flux $\Phi$ is show in Fig.~\ref{fig3}%
(a). At time $\Omega _{ab} t=1$, we have $T_{ab}(t)> T_{ba}(t)$ for synthetic
magnetic flux $0<\Phi<\pi$; in the contrast, we have $T_{ab}(t)< T_{ba}(t)$
for synthetic magnetic flux $-\pi<\Phi<0$. As shown in Fig.~\ref{fig3}(b),
under the resonant condition $\Delta _{cb}=\Delta_{ca}=\Delta _{ab} =0$, the
optimal isolation $I(t)$ is obtained with synthetic magnetic flux $\Phi =\pm
\pi /2$.

\emph{The implementation of cyclic three-level transition.---}
To realize a cyclic three-level atom, one ingredient is breaking the symmetry of
the potential of the atom. The cyclic three-level transition has been proposed and observed in chiral molecules~\cite{KralPRL01,KralPRL03,YLiPRL07,PattersonPRL13,PattersonNat13,EibenbergerPRL17,CYePRA18}. In addition, the potential of the atom also can be broken by applying
an external magnetic field. We can consider a qubit circuit composed of a superconducting loop with three Josephson junctions~\cite{YXLiuPRL05,MooijSci99} that encloses an
applied magnetic flux $\Phi_{e}=f\Phi_{0}$ ($\Phi_{0}\equiv h/2e$ is the superconducting
flux quantum, where $h$ is Planck's constant and $f\equiv \Phi_{e}/\Phi_{0}$ is the reduced magnetic flux). When the reduced magnetic flux $f$ is a half-integer, the potential of the artificial atom is symmetric, and the
interaction Hamiltonian has odd parity. However, when $f$ is not a half-integer, the symmetry of the potential is broken and the interaction Hamiltonian does not have well defined parity. In this case, transitions can
occur between any two levels.

Alternatively, cyclic transitions in three-level atom can be realized by a
single nitrogen-vacancy (NV) center embedded in a mechanical resonator~\cite%
{BarfussArx18}. Three eigenstates ($%
\left\vert 0\right\rangle $ and $\left\vert \pm 1\right\rangle $) of the
spin operator along the NV's symmetry axis $z$ (i.e., $S_{z}\left\vert
m\right\rangle=m\left\vert m\right\rangle$) are selected as a three-level
atom~\cite{MazeNJP11,DobrovitskiARCMP13}. The two degenerate levels $%
\left\vert \pm 1\right\rangle $ can be split by applying an external
magnetic field along $z$. We can use microwave magnetic fields to drive the
transitions between $\left\vert 0\right\rangle $ and $\left\vert \pm
1\right\rangle $; the magnetic dipole-forbidden transition $\left\vert +
1\right\rangle \leftrightarrow \left\vert - 1\right\rangle$ can be driven by
a time-varying strain field through the mechanical resonator~\cite%
{MacQuarriePRL13,BarfussNPy15}.

\begin{figure}[tbp]
\includegraphics[bb=3 311 590 664, width=8.5 cm, clip]{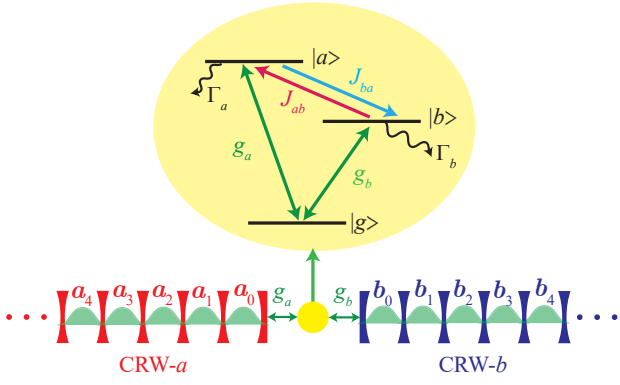}
\caption{(Color online) Schematic of two 1D semi-infinite CRWs connected by
a three-level atom characterized by $\left\vert a\right\rangle $, $%
\left\vert b\right\rangle $, and $\left\vert g\right\rangle $. CRW-$a$ (CRW-$%
b$) couples to the three-level atom through the transition $\left\vert
a\right\rangle \leftrightarrow \left\vert g\right\rangle$ ($\left\vert
b\right\rangle \leftrightarrow \left\vert g\right\rangle$) with strength $%
g_{a}$ ($g_{b}$).}
\label{fig4}
\end{figure}

\emph{Single-photon nonreciprocal transport.---} As an important
application, we will discuss how to realize a single-photon nonreciprocal
transport between two one dimensional (1D) semi-infinite coupled-resonator
waveguides (CRWs) by nonreciprocal transition effect. We assume that two 1D
semi-infinite CRWs, with creation operators $a_{j}^{\dag }$ and $b_{j}^{\dag
}$ and frequency $\omega_{w,a}$ ($\omega_{w,b}$) for the $j$th cavity modes,
are coupled by a $\nabla $-type three-level atom ($\left\vert a\right\rangle
$, $\left\vert b\right\rangle $, and $\left\vert g\right\rangle $) with
nonreciprocal transition $\left\vert a\right\rangle \leftrightarrow
\left\vert b\right\rangle $, as shown in Fig.~\ref{fig4}. Here $g_{a}$ ($%
g_{b}$) is the coupling strength between CRW-$a$ (CRW-$b$) and the
transition $\left\vert a\right\rangle \leftrightarrow \left\vert
g\right\rangle$ ($\left\vert b\right\rangle \leftrightarrow \left\vert
g\right\rangle$) with frequency $\omega_{ag}$ ($\omega_{bg}$). The system
can be described by the total Hamiltonian under the rotating wave
approximation $H_{\mathrm{tot}}=\sum_{l=a,b}H_{l}+\widetilde{H}_{\mathrm{eff}%
}+H_{\mathrm{int}}$. Here, in the rotating reference frame with respect to $%
H_{\mathrm{rot}}=\omega _{ag}\left( \sum_{j}a_{j}^{\dag }a_{j}+\left\vert
a\right\rangle \left\langle a\right\vert \right) +\omega _{bg}\left(
\sum_{j}b_{j}^{\dag }b_{j}+\left\vert b\right\rangle \left\langle
b\right\vert \right) $, the Hamiltonian $H_{l}$ for the CRW-$l$ is given by%
\begin{equation}
H_{l}=\Delta _{l}\sum_{j=0}^{+\infty }l_{j}^{\dag }l_{j}-\xi
_{l}\sum_{j=0}^{+\infty }\left( l_{j}^{\dag }l_{j+1}+\mathrm{H.c.}\right)
\end{equation}%
with homogeneous intercavity coupling constants $\xi_{l}$ and cavity-atom
detunings $\Delta _{l}=\omega_{w,l}-\omega _{lg}$ ($l=a,b$); the effective
Hamiltonian $\widetilde{H}_{\mathrm{eff}}$ for the $\nabla $-type
three-level atom with nonreciprocal transition $\left\vert a\right\rangle
\leftrightarrow \left\vert b\right\rangle $ is obtained from Eq.~(\ref{eq4})
with $\Delta_a=\Delta_b=0$ as
\begin{eqnarray}  \label{eq28}
\widetilde{H}_{\mathrm{eff}} &=&J_{ab}\left\vert a\right\rangle \left\langle
b\right\vert +J_{ba}\left\vert b\right\rangle \left\langle a\right\vert
\nonumber \\
&&-i\Gamma _{a}\left\vert a\right\rangle \left\langle a\right\vert -i\Gamma
_{b}\left\vert b\right\rangle \left\langle b\right\vert,
\end{eqnarray}%
and the interaction Hamiltonian $H_{\mathrm{int}}$ between the $0$th cavity
modes and the three-level atom is described by%
\begin{eqnarray}
H_{\mathrm{int}} &=&g_{a}a_{0}\left\vert a\right\rangle \left\langle
g\right\vert +g_{b}b_{0}\left\vert b\right\rangle \left\langle g\right\vert
\nonumber \\
&&+g_{a}a_{0}^{\dag }\left\vert g\right\rangle \left\langle a\right\vert
+g_{b}b_{0}^{\dag }\left\vert g\right\rangle \left\langle b\right\vert.
\end{eqnarray}

The efficiency for nonreciprocity transport can be described by the the
scattering flow~\cite{ZHWangPRA14,XWXuPRA17a,XWXuPRA17b,LZhouPRL13} $%
I_{l^{\prime }l}$ for a single photon from CRW-$l$ to CRW-$l^{\prime }$ ($%
l=a,b$). The detailed calculations of the scattering flow $I_{l^{\prime }l}$
can be found in the supplementary material~\cite{SM}. Nonreciprocal
single-photon transport appears when $I_{ba}\neq I_{ab}$, which implies that
the scattering flow from CRW-$a$ to CRW-$b$ is not equal to the scattering
flow from CRW-$b$ to CRW-$a$.

\begin{figure}[tbp]
\includegraphics[bb=83 245 531 592, width=8.5 cm, clip]{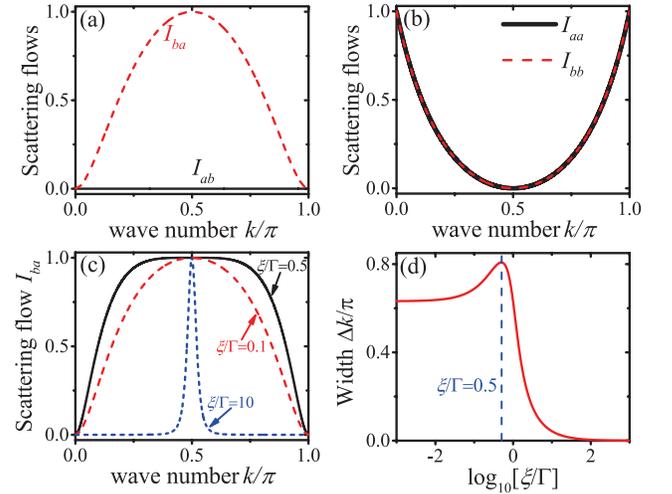}
\caption{(Color online) (a) Scattering flows $I_{ab}$ (black solid curve)
and $I_{ba}$ (red dashed curve), (b) $I_{aa}$ (black solid curve) and $%
I_{bb} $ (red dashed curve), are plotted as functions of the wave number $k/%
\protect\pi$ for $\protect\xi/\Gamma=0.1$. (c) Scattering flow $I_{ab}$ is
plotted as a function of the wave number $k/\protect\pi$ for different $%
\protect\xi/\Gamma$. (d) The width of the wave number $\Delta k$ for
single-photon nonreciprocity is plotted as a function of $\log_{10}[\protect%
\xi/\Gamma]$ given in Eq.~(\protect\ref{eq33}). The other parameters are $%
J_{ba}=2\Gamma$, $J_{ab}=0$, $\protect\xi=\Gamma$, $\Delta _{a}=\Delta
_{b}=0 $, $g^{2}=\Gamma \protect\xi$, $\protect\phi=\protect\pi/2$.}
\label{fig5}
\end{figure}

First of all, let us find the optimal conditions for perfect single-photon
nonreciprocity, i.e., $I_{ab}=0$ and $I_{ba}=1$, analytically. For
simplicity, we assume that the two semi-infinite CRWs have the same
parameters, i.e., $\xi \equiv \xi _{a}=\xi _{b}$, $k\equiv k_{a}=k_{b}$, $%
g\equiv g_{a}=g_{b}$, and they are coupled to the atom resonantly with $%
\Delta _{a}=\Delta _{b}=0$ and $\Gamma \equiv \Gamma _{a}=\Gamma _{b}$.
$I_{ab}=0$ can be obtained by setting $J_{ab}=0$. Through a simple derivation~\cite{SM}, the condition for $I_{ba}=1$ is $|\sin k | =1$, i.e., $k=\pi/2$ ($0<k<\pi$), in the case that $\vert J_{ba}\vert =2\Gamma$ and $g^{2}=\Gamma \xi$. This fits the numerical simulation very well, as shown in Fig.~\ref{fig5}(a) and \ref{fig5}(b).

Now let us discuss the width of the wave number for single-photon
nonreciprocity~\cite{SM}. We define the width of the wave number $\Delta k$ for
single-photon nonreciprocity as the full width at half maximum (FWHM) by setting $I_{ba}=1/2$
for $k=k_{\mathrm{half}}\in [0,\pi/2)$,
\begin{equation}  \label{eq33}
\Delta k\equiv\pi -2k_{\mathrm{half}}.
\end{equation}
Under the conditions $\vert J_{ba}\vert =2\Gamma$ and $g^{2}=\Gamma \xi$, there is a maximum FWHM for single-photon nonreciprocity at $\xi =\Gamma /2$, and the maximum FWHM $\Delta k _{\mathrm{max}}\approx 0.81\pi$ is
obtained with $k_{\mathrm{half}}=\sin^{-1} (2\sqrt{2}-1-2\sqrt{2-\sqrt{2}})$%
, in excellent agreement with Fig.~\ref{fig5}(c) and \ref{fig5}(d).

\emph{Conclusion.---} In summary, we have shown theoretically that nonreciprocal transition
can be observed between two nondegenerate energy levels.
A general method has been presented to
realize nonreciprocal transition between two nondegenerate energy levels based on a
combination of synthetic magnetism and reservoir engineering. As a simple
example, we explicitly show an implementation involving an auxiliary energy
level, i.e., a cyclic three-level atom system. A single-photon nonreciprocal
transporter has been proposed by the nonreciprocal transition effect.
The atom-mediated nonreciprocal devices based on the nonreciprocal transition are suitable for applications in building hybrid quantum networks.

The generic method for realizing nonreciprocal transition can also be
applied in quantum acoustodynamical systems~\cite%
{OConnellNat10,YChuSci17,ManentiNC17} to design nonreciprocal phonon
devices. Furthermore, the nonreciprocal transition can also be implemented in a
wide range of physical systems, such as a four-level atom system~\cite%
{RipkaSci18}, two qubits in a one-dimension waveguide~\cite{HamannPRL18},
and even qubit arrays~\cite{JohnsonNat11}. The nonreciprocal transition can
be extended to explore quantum nonreciprocal physics~\cite%
{RHuangPRL18,XuXWArx18,BLiPR19} and topological phases~\cite{HasanRMP10} in a single
multi-level atom or qubit array. Besides these, the nonreciprocal transition effect can also be used to avoid the echo formation in quantum memory and quantum measurements.

\begin{acknowledgements}
\emph{Acknowledgements.}---
X.-W.X. was supported by the National Natural Science Foundation of China
(NSFC) under Grant No.~11604096, and the Key Program of Natural Science Foundation of Jiangxi Province, China under Grant No.~20192ACB21002.
Y.J.Z. is supported by NSFC under Grants No.~11904013 and No.~11847165.
A.X.C. is supported by NSFC under Grant No.~11775190. Y.X.L. is supported by the National Basic
Research Program of China (973 Program) under Grant No.~2014CB921401, the
Tsinghua University Initiative Scientific Research Program, and the Tsinghua
National Laboratory for Information Science and Technology (TNList)
Cross-discipline Foundation.
\end{acknowledgements}

\bibliographystyle{apsrev}
\bibliography{ref}

\clearpage

\onecolumngrid

\setcounter{equation}{0} \setcounter{figure}{0} \setcounter{table}{0} %
\setcounter{page}{1} \makeatletter
\renewcommand{\theequation}{S\arabic{equation}} \renewcommand{\thefigure}{S%
\arabic{figure}} \renewcommand{\bibnumfmt}[1]{[S#1]} \renewcommand{%
\citenumfont}[1]{S#1} \renewcommand\thesection{S\arabic{section}}

\begin{center}
{\large \textbf{Supplementary Material for ``Nonreciprocal transition
between two nondegenerate energy levels''}}

Xun-Wei Xu$^{1}$, Yan-Jun Zhao$^{2}$, Hui Wang$^3$, Ai-Xi Chen$^{4,1}$, and
Yu-xi Liu$^{5,6}$
\end{center}

\begin{minipage}[]{16cm}
\small{\it

\centering $^{1}$ Department of Applied Physics, East China Jiaotong University, Nanchang, 330013, China  \\
\centering $^{2}$ Faculty of Information Technology, College of Microelectronics,  \\
\centering Beijing University of Technology, Beijing 100124, China,  \\
\centering $^{3}$ Center for Emergent Matter Science (CEMS), RIKEN, Wako, Saitama 351-0198, Japan  \\
\centering $^{4}$ Department of Physics, Zhejiang Sci-Tech University, Hangzhou, 310018, China  \\
\centering $^{5}$ Institute of Microelectronics, Tsinghua University, Beijing 100084, China \\
\centering $^{6}$ Beijing National Research Center for Information Science and Technology (BNRist), Beijing 100084, China \\
}

\end{minipage}

\vspace{8mm}

\section{Adiabatic elimination}

We consider a cyclic three-level atom ($\left\vert a\right\rangle $, $%
\left\vert b\right\rangle $, and $\left\vert c\right\rangle $) driven by
three classical coherent fields (at rates $\Omega _{ij}$, phases $\phi _{ij}$%
, frequencies $\nu _{ij}$, with $i,j=a,b,c$ and $\nu _{cb}=\nu _{ab}+\nu
_{ca}$) that is described by a Hamiltonian given by
\begin{eqnarray}
\widetilde{H} &=&\left( \omega _{ab}-i\gamma _{a}\right) \left\vert
a\right\rangle \left\langle a\right\vert -i\gamma _{b}\left\vert
b\right\rangle \left\langle b\right\vert +\left( \omega _{cb}-i\gamma
_{c}\right) \left\vert c\right\rangle \left\langle c\right\vert  \nonumber
\label{eq6} \\
&&+\left( \Omega _{ab}e^{i\phi _{ab}}e^{-i\nu _{ab}t}\left\vert
a\right\rangle \left\langle b\right\vert +\Omega _{cb}e^{-i\phi
_{cb}}e^{-i\nu _{cb}t}\left\vert c\right\rangle \left\langle b\right\vert
+\Omega _{ca}e^{i\phi _{ca}}e^{-i\nu _{ca}t}\left\vert c\right\rangle
\left\langle a\right\vert +\mathrm{H.c.}\right),
\end{eqnarray}%
where $\omega _{ij}$ is the frequency difference between levels $\left\vert
i\right\rangle $ and $\left\vert j\right\rangle $, and the three levels can
decay to the other levels with the decay rates $\gamma _{i}$ $\left(
i=a,b,c\right) $.

In the rotating frame respect to the operator%
\[
W=e^{-i\left( \nu _{ab}\left\vert a\right\rangle \left\langle a\right\vert
+\nu _{cb}\left\vert c\right\rangle \left\langle c\right\vert \right) t}
\]%
we have%
\begin{eqnarray*}
H &=&U^{\dag }\widetilde{H}U+i\frac{dW^{\dag }}{dt}W \\
&=&\left( \Delta _{ab}-i\gamma _{a}\right) \left\vert a\right\rangle
\left\langle a\right\vert -i\gamma _{b}\left\vert b\right\rangle
\left\langle b\right\vert +\left( \Delta _{cb}-i\gamma _{c}\right)
\left\vert c\right\rangle \left\langle c\right\vert \\
&&+\Omega _{ab}e^{i\phi _{ab}}\left\vert a\right\rangle \left\langle
b\right\vert +\Omega _{cb}e^{-i\phi _{cb}}\left\vert c\right\rangle
\left\langle b\right\vert +\Omega _{ca}e^{i\phi _{ca}}\left\vert
c\right\rangle \left\langle a\right\vert +\mathrm{H.c.},
\end{eqnarray*}%
with the detuning $\Delta _{ij}\equiv \omega _{ij}-\nu _{ij}$ $\left(
i,j=a,b,c\right) $. By local redefinition of the eigenstates, i.e., $%
e^{-i\phi _{cb}}\left\langle b\right\vert \rightarrow \left\langle
b\right\vert $ and $e^{-i\phi _{ca}}\left\vert a\right\rangle \rightarrow
\left\vert a\right\rangle $, the Hamiltonian can be rewritten as%
\begin{eqnarray}
H &=&\left( \Delta _{ab}-i\gamma _{a}\right) \left\vert a\right\rangle
\left\langle a\right\vert -i\gamma _{b}\left\vert b\right\rangle
\left\langle b\right\vert +\left( \Delta _{cb}-i\gamma _{c}\right)
\left\vert c\right\rangle \left\langle c\right\vert  \nonumber \\
&&+\left( \Omega _{ab}e^{i\Phi }\left\vert a\right\rangle \left\langle
b\right\vert +\Omega _{cb}\left\vert c\right\rangle \left\langle
b\right\vert +\Omega _{ca}\left\vert c\right\rangle \left\langle
a\right\vert +\mathrm{H.c.}\right) ,
\end{eqnarray}%
with the synthetic magnetic flux%
\[
\Phi \equiv\phi _{ab}+\phi _{cb}+\phi _{ca}.
\]

Now we consider the transitions between the three levels. The state vector
for this three levels at time $t$ can be written as%
\begin{equation}
\left\vert \psi \right\rangle =A\left( t\right) \left\vert a\right\rangle
+B\left( t\right) \left\vert b\right\rangle +C\left( t\right) \left\vert
c\right\rangle .
\end{equation}%
The coefficients $|A(t)|^{2}$, $|B(t)|^{2}$, and $|C(t)|^{2}$ denote
occupying probabilities in states $\left\vert a\right\rangle $, $\left\vert
b\right\rangle $ and $\left\vert c\right\rangle $, respectively. Then the
dynamical behaviors for the coefficients can be obtained by the Schr\"{o}%
dinger equation, i.e., $i\dot{\left\vert \psi \right\rangle }=H\left\vert
\psi \right\rangle $, given by
\begin{equation}
\dot{A}\left( t\right) =\left( -i\Delta _{ab}-\gamma _{a}\right) A\left(
t\right) -i\Omega _{ab}e^{i\Phi }B\left( t\right) -i\Omega _{ca}C\left(
t\right) ,  \label{eqS2}
\end{equation}%
\begin{equation}
\dot{B}\left( t\right) =-\gamma _{b}B\left( t\right) -i\Omega _{ab}e^{-i\Phi
}A\left( t\right) -i\Omega _{cb}C\left( t\right) ,  \label{eqS3}
\end{equation}%
\begin{equation}
\dot{C}\left( t\right) =\left( -i\Delta _{cb}-\gamma _{c}\right) C\left(
t\right) -i\Omega _{ca}A\left( t\right) -i\Omega _{cb}B\left( t\right) .
\label{eqS4}
\end{equation}%
Under the assumption that the decay of the state $\left\vert c\right\rangle $
is much faster than decay of the states $\left\vert a\right\rangle $ and $%
\left\vert b\right\rangle $, i.e., $\gamma _{c}\gg \left\{ \gamma
_{a},\gamma _{b}\right\} $, we can adiabatically eliminate the level $%
\left\vert c\right\rangle $ with $\dot{C}\left( t\right) =0$ as%
\begin{equation}
C\left( t\right) =\frac{-i\Omega _{ca}}{\gamma _{c}+i\Delta _{cb}}A\left(
t\right) -\frac{i\Omega _{cb}}{\gamma _{c}+i\Delta _{cb}}B\left( t\right) .
\label{eqS5}
\end{equation}%
By substituting Eq.~(\ref{eqS5}) into Eqs.~(\ref{eqS2}) and (\ref{eqS3}),
then the dynamical equations of $A\left( t\right) $ and $B\left( t\right) $
become%
\begin{eqnarray}
\dot{A}\left( t\right) &=&-\left[ i\left( \Delta _{ab}-\frac{\Omega
_{ca}^{2}\Delta _{cb}}{\gamma _{c}^{2}+\Delta _{cb}^{2}}\right) +\left(
\gamma _{a}+\frac{\Omega _{ca}^{2}\gamma _{c}}{\gamma _{c}^{2}+\Delta
_{cb}^{2}}\right) \right] A\left( t\right)  \nonumber \\
&&-\left[ i\Omega _{ab}e^{i\Phi }+\frac{\Omega _{ca}\Omega _{cb}\left(
\gamma _{c}-i\Delta _{cb}\right) }{\gamma _{c}^{2}+\Delta _{cb}^{2}}\right]
B\left( t\right) ,  \label{eqS6}
\end{eqnarray}%
\begin{eqnarray}
\dot{B}\left( t\right) &=&-\left[ -i\frac{\Omega _{cb}^{2}\Delta _{cb}}{%
\gamma _{c}^{2}+\Delta _{cb}^{2}}+\left( \gamma _{b}+\frac{\Omega
_{cb}^{2}\gamma _{c}}{\gamma _{c}^{2}+\Delta _{cb}^{2}}\right) \right]
B\left( t\right)  \nonumber \\
&&-\left[ i\Omega _{ab}e^{-i\Phi }+\frac{\Omega _{ca}\Omega _{cb}\left(
\gamma _{c}-i\Delta _{cb}\right) }{\gamma _{c}^{2}+\Delta _{cb}^{2}}\right]
A\left( t\right) ,  \label{eqS7}
\end{eqnarray}%
Physically, the effective Hamiltonian representing the Schr\"{o}dinger
evolution in Eqs.~(\ref{eqS6}) and (\ref{eqS7}) is given by
\begin{eqnarray}
H_{\mathrm{eff}} &=&\left( \Delta _{a}-i\Gamma _{a}\right) \left\vert
a\right\rangle \left\langle a\right\vert +\left( \Delta _{b}-i\Gamma
_{b}\right) \left\vert b\right\rangle \left\langle b\right\vert  \nonumber
\label{eqS8} \\
&&+J_{ab}\left\vert a\right\rangle \left\langle b\right\vert
+J_{ba}\left\vert b\right\rangle \left\langle a\right\vert
\end{eqnarray}%
with effective detunings
\[
\Delta _{a}\equiv \Delta _{ab}-\frac{\Omega _{ca}^{2}\Delta _{cb}}{\gamma
_{c}^{2}+\Delta _{cb}^{2}},
\]%
\[
\Delta _{b}\equiv -\frac{\Omega _{cb}^{2}\Delta _{cb}}{\gamma
_{c}^{2}+\Delta _{cb}^{2}},
\]%
effective decay rates%
\[
\Gamma _{a}\equiv \left( \gamma _{a}+\frac{\Omega _{ca}^{2}\gamma _{c}}{%
\gamma _{c}^{2}+\Delta _{cb}^{2}}\right) ,
\]%
\[
\Gamma _{b}\equiv \left( \gamma _{b}+\frac{\Omega _{cb}^{2}\gamma _{c}}{%
\gamma _{c}^{2}+\Delta _{cb}^{2}}\right) ,
\]%
and effective coupling coefficients
\[
J_{ab}\equiv \Omega _{ab}e^{i\Phi }-i\frac{\Omega _{ca}\Omega _{cb}\left(
\gamma _{c}-i\Delta _{cb}\right) }{\gamma _{c}^{2}+\Delta _{cb}^{2}},
\]%
\[
J_{ba}\equiv \Omega _{ab}e^{-i\Phi }-i\frac{\Omega _{ca}\Omega _{cb}\left(
\gamma _{c}-i\Delta _{cb}\right) }{\gamma _{c}^{2}+\Delta _{cb}^{2}}.
\]%
The effective coupling coefficients $J_{ab}$ and $J_{ba}$ are dependent on
the synthetic magnetic flux $\Phi $. Under the resonant condition $\Delta
_{cb}=0$, we have $|J_{ab}|<|J_{ba}|$ for $0<\Phi <\pi $; conversely, we
have $|J_{ab}|>|J_{ba}|$ as $-\pi <\Phi <0$. The perfect nonreciprocal
transition, i.e, $J_{ab}=0$ and $J_{ba}\neq 0$ (or $J_{ba}=0$ and $%
J_{ab}\neq 0$), is obtained when $\Phi =\pi /2$ (or $\Phi =-\pi /2$) with $%
\Omega _{ab}=\Omega _{ca}\Omega _{cb}/\gamma _{c}$.

\section{Scattering flow}

To study the nonreciprocal single-photon transport, we discuss the
scattering of a single photon in the system with the total Hamiltonian%
\begin{equation}
H_{\mathrm{tot}}=\sum_{l=a,b}H_{l}+\widetilde{H}_{\mathrm{eff}}+H_{\mathrm{%
int}}.
\end{equation}%
Here, in the rotating reference frame with respect to
\[
H_{\mathrm{rot}}=\omega _{ag}\left( \sum_{j}a_{j}^{\dag }a_{j}+\left\vert
a\right\rangle \left\langle a\right\vert \right) +\omega _{bg}\left(
\sum_{j}b_{j}^{\dag }b_{j}+\left\vert b\right\rangle \left\langle
b\right\vert \right) ,
\]
the Hamiltonian $H_{l}$ for the CRW-$l$ is given by%
\[
H_{l}=\Delta _{l}\sum_{j=0}^{+\infty }l_{j}^{\dag }l_{j}-\xi
_{l}\sum_{j=0}^{+\infty }\left( l_{j}^{\dag }l_{j+1}+\mathrm{H.c.}\right)
\]%
with homogeneous intercavity coupling constants $\xi _{l}$ and cavity-atom
detunings $\Delta _{l}=\omega _{w,l}-\omega _{lg}$ ($l=a,b$); the effective
Hamiltonian $\widetilde{H}_{\mathrm{eff}}$ for the $\nabla $-type
three-level atom with nonreciprocal transition $\left\vert a\right\rangle
\leftrightarrow \left\vert b\right\rangle $ with $\Delta _{a}=\Delta _{b}=0$
is given by
\begin{eqnarray*}
\widetilde{H}_{\mathrm{eff}} &=&J_{ab}\left\vert a\right\rangle \left\langle
b\right\vert +J_{ba}\left\vert b\right\rangle \left\langle a\right\vert \\
&&-i\Gamma _{a}\left\vert a\right\rangle \left\langle a\right\vert -i\Gamma
_{b}\left\vert b\right\rangle \left\langle b\right\vert ,
\end{eqnarray*}%
and the interaction Hamiltonian $H_{\mathrm{int}}$ between the $0$th cavity
modes and the three-level system is described by%
\begin{eqnarray*}
H_{\mathrm{int}} &=&g_{a}a_{0}\left\vert a\right\rangle \left\langle
g\right\vert +g_{b}b_{0}\left\vert b\right\rangle \left\langle g\right\vert
\\
&&+g_{a}a_{0}^{\dag }\left\vert g\right\rangle \left\langle a\right\vert
+g_{b}b_{0}^{\dag }\left\vert g\right\rangle \left\langle b\right\vert .
\end{eqnarray*}

As the total number of photons in the system is a conserved quantity
(without dissipation), we consider the stationary eigenstate of a single
photon in the system as
\begin{equation}
\left\vert E\right\rangle =\sum_{j=0}^{+\infty }\left[ u_{a}\left( j\right)
a_{j}^{\dag }+u_{b}\left( j\right) b_{j}^{\dag }\right] \left\vert
g,0\right\rangle +A\left\vert a,0\right\rangle +B\left\vert b,0\right\rangle
,  \label{eq34}
\end{equation}%
where $\left\vert 0\right\rangle $ indicates the vacuum state of the 1D
semi-infinite CRWs, $u_{l}\left( j\right) $ denotes the probability
amplitude in the state with a single photon in the $j$th cavity of the CRW-$%
l $, and $A$ ($B$) denotes the probability amplitude in the atom state $%
\left\vert a\right\rangle $ ($\left\vert b\right\rangle $). The dispersion
relation of the semi-infinite CRW-$l$ in the rotating reference frame is
given by~\cite{SMLZhouPRL13}%
\[
E=\Delta _{l}-2\xi _{l}\cos k_{l},\quad 0<k_{l}<\pi ,
\]%
where $E_{l}$ is the energy and $k_{l}$ is the wave number of the single
photon in the CRW-$l$. Without loss of generality, we assume that $\xi
_{l}>0 $. Substituting the stationary eigenstate $\left\vert E\right\rangle $
in Eq.~(\ref{eq34}) and the total Hamiltonian $H_{\mathrm{tot}}$ into the
eigenequation $H_{\mathrm{tot}}\left\vert E\right\rangle =E\left\vert
E\right\rangle $, we can obtain the coupled equations for the probability
amplitudes as%
\begin{equation}
\Delta _{a}u_{a}\left( 0\right) -\xi _{a}u_{a}\left( 1\right)
+g_{a}A=Eu_{a}\left( 0\right)  \label{eq35}
\end{equation}%
\begin{equation}
\Delta _{b}u_{b}\left( 0\right) -\xi _{b}u_{b}\left( 1\right)
+g_{b}B=Eu_{b}\left( 0\right)
\end{equation}%
\begin{equation}
-i\Gamma _{a}A+g_{a}u_{a}\left( 0\right) +J_{ab}B=EA
\end{equation}%
\begin{equation}
-i\Gamma _{b}B+g_{b}u_{b}\left( 0\right) +J_{ba}A=EB
\end{equation}%
\begin{equation}
\Delta _{l}u_{l}\left( j\right) -\xi _{l}\left[ u_{l}\left( j+1\right)
+u_{l}\left( j-1\right) \right] =Eu_{l}\left( j\right)  \label{eq39}
\end{equation}%
with $j>0$ and $l=a,b$.

If a single photon with energy $E$ is incident from the infinity side of CRW-%
$l$, the $\nabla$-type three-level atom will result in photon scattering
between different CRWs or photon absorbtion by the dissipative of the atom.
The general expressions of the probability amplitudes in the CRWs ($j\geq 0$%
) are given by
\begin{equation}
u_{l}\left( j\right) =e^{-ik_{l}j}+s_{ll}e^{ik_{l}j},  \label{eq40}
\end{equation}%
\begin{equation}
u_{l^{\prime }}\left( j\right) =s_{l^{\prime }l}e^{ik_{l^{\prime }}j},
\label{eq41}
\end{equation}%
where $s_{l^{\prime }l}$ denotes the single-photon scattering amplitude from
CRW-$l$ to CRW-$l^{\prime }$ ($l,l^{\prime }=a,b$). Substituting Eqs.~(\ref%
{eq40}) and (\ref{eq41}) into Eqs.~(\ref{eq35})-(\ref{eq39}), then we obtain
the scattering matrix as%
\begin{equation}
S=\left(
\begin{array}{cc}
s_{aa} & s_{ab} \\
s_{ba} & s_{bb}%
\end{array}%
\right) ,
\end{equation}%
where
\begin{equation}
s_{aa}=D^{-1}\left[ J_{ab}^{\prime }J_{ba}^{\prime }-\left( \xi
_{a}e^{ik_{a}}+\overline{\Delta }_{a}\right) \left( \xi _{b}e^{-ik_{b}}+%
\overline{\Delta }_{b}\right) \right] ,
\end{equation}%
\begin{equation}
s_{ba}=i2D^{-1}J_{ba}^{\prime }\xi _{a}\sin k_{a},
\end{equation}%
\begin{equation}
s_{ab}=i2D^{-1}J_{ab}^{\prime }\xi _{b}\sin k_{b},
\end{equation}%
\begin{equation}
s_{bb}=D^{-1}\left[ J_{ab}^{\prime }J_{ba}^{\prime }-\left( \xi
_{a}e^{-ik_{a}}+\overline{\Delta }_{a}\right) \left( \xi _{b}e^{ik_{b}}+%
\overline{\Delta }_{b}\right) \right] ,
\end{equation}%
\begin{equation}
D=\left( \xi _{a}e^{-ik_{a}}+\overline{\Delta }_{a}\right) \left( \xi
_{b}e^{-ik_{b}}+\overline{\Delta }_{b}\right) -J_{ab}^{\prime
}J_{ba}^{\prime }
\end{equation}%
with the effective coupling strengths $J^{\prime }_{ll^{\prime }}$ and
frequency shifts $\overline{\Delta } _{l}$ induced by the $\nabla$-type
three-level atom defined by
\begin{equation}
J_{ab}^{\prime }\equiv \frac{J_{ab}g_{a}g_{b}}{\left( E+i\Gamma _{a}\right)
\left( E+i\Gamma _{b}\right) -J_{ba}J_{ab}},
\end{equation}%
\begin{equation}
J_{ba}^{\prime }\equiv \frac{J_{ba}g_{a}g_{b}}{\left( E+i\Gamma _{a}\right)
\left( E+i\Gamma _{b}\right) -J_{ba}J_{ab}},
\end{equation}%
\begin{equation}
\overline{\Delta }_{a} \equiv \frac{\left( E+i\Gamma _{b}\right) g_{a}^{2}}{%
\left( E+i\Gamma _{a}\right) \left( E+i\Gamma _{b}\right) -J_{ba}J_{ab}},
\end{equation}%
\begin{equation}
\overline{\Delta }_{b} \equiv \frac{\left( E+i\Gamma _{a}\right) g_{b}^{2}}{%
\left( E+i\Gamma _{a}\right) \left( E+i\Gamma _{b}\right) -J_{ba}J_{ab}}.
\end{equation}

To quantify the efficiency for nonreciprocity transport, we define the the
scattering flow~\cite{SMZHWangPRA14,SMXWXuPRA17a,SMXWXuPRA17b} of a single
photon from CRW-$l$ to CRW-$l^{\prime }$, as
\begin{equation}
I_{l^{\prime }l}\equiv |s_{l^{\prime }l}|^{2}\frac{\xi _{l^{\prime }}\sin
k_{l^{\prime }}}{\xi _{l}\sin k_{l}},
\end{equation}%
where $\xi _{l}\sin k_{l}$ ($\xi _{l^{\prime }}\sin k_{l^{\prime }}$) is the
group velocity in the CRW-$l$ (CRW-$l^{\prime }$).

\section{Perfect single-photon nonreciprocity}

In this section, we will derive the conditions for perfect nonreciprocal
single-photon transport, i.e., $I_{ab}=0$ and $I_{ba}=1$, analytically. For
simplicity, we assume that the two semi-infinite CRWs have the same
parameters, i.e., $\xi \equiv \xi _{a}=\xi _{b}$, $k\equiv k_{a}=k_{b}$, $%
g\equiv g_{a}=g_{b}$, and they are coupled to the atom resonantly with $%
\Delta _{a}=\Delta _{b}=0$ and $\Gamma \equiv \Gamma _{a}=\Gamma _{b}$. $%
I_{ab}=0$ can be obtained by setting $J_{ab}=0$ or $J_{ab}^{\prime }=0$. In
this case, we have%
\begin{eqnarray*}
I_{ba} &=&\left\vert s_{ba}\right\vert ^{2}=\left\vert \frac{%
2iJ_{ba}^{\prime }\xi _{a}\sin k_{a}}{\left( \xi _{a}e^{-ik_{a}}+\overline{%
\Delta }_{a}\right) \left( \xi _{b}e^{-ik_{b}}+\overline{\Delta }_{b}\right)
}\right\vert ^{2} \\
&=&\left\vert \frac{2J_{ba}g_{a}g_{b}\xi _{a}\sin k_{a}}{\left[ \xi
_{a}e^{-ik_{a}}\left( E+i\Gamma _{a}\right) +g_{a}^{2}\right] \left[ \xi
_{b}e^{-ik_{b}}\left( E+i\Gamma _{b}\right) +g_{b}^{2}\right] }\right\vert
^{2} \\
&=&\left\vert \frac{2J_{ba}g^{2}\xi \sin k}{\left[ \xi e^{-ik}\left( -2\xi
\cos k+i\Gamma \right) +g^{2}\right] ^{2}}\right\vert ^{2}.
\end{eqnarray*}%
So the condition for $I_{ba}=1$ is%
\[
2\left\vert J_{ba}\right\vert g^{2}\xi \left\vert \sin k\right\vert
=\left\vert \xi e^{-ik}\left( -2\xi \cos k+i\Gamma \right) +g^{2}\right\vert
^{2}
\]%
or%
\[
\left\vert \sin k\right\vert =\frac{\left( \left\vert J_{ba}\right\vert
-\Gamma \right) g^{2}\pm \sqrt{\left( \left\vert J_{ba}\right\vert -\Gamma
\right) ^{2}g^{4}-4\left( g^{2}-\xi ^{2}\right) \left[ 4\left( \xi
^{2}-g^{2}\right) \xi ^{2}+\Gamma ^{2}\xi ^{2}+g^{4}\right] }}{4\left(
g^{2}-\xi ^{2}\right) \xi }.
\]

As a simple example, the maximum scattering flow $I_{ba}=1$ can be obtained at the maximum group velocity
\[
\left\vert \sin k\right\vert =1
\]%
with
\[
\left\vert J_{ba}\right\vert =\frac{\left( g^{2}+\Gamma \xi \right) ^{2}}{%
2g^{2}\xi }.
\]%
Furthermore, if
\[
g^{2}=\Gamma \xi,
\]%
we have%
\[
\left\vert J_{ba}\right\vert =2\Gamma.
\]

\section{Maximum full width at half maximum}

Now, we will derive the maximum full width at half maximum (FWHM) for perfect
nonreciprocal single-photon transport. The half maximum of the scattering
flow $I_{ba}$ is given by%
\[
I_{ba}=\left\vert s_{ba}\right\vert ^{2}=\left\vert \frac{2J_{ba}g^{2}\xi
\sin k_{\mathrm{half}}}{\left[ \xi e^{-ik}\left( -2\xi \cos k_{\mathrm{half}%
}+i\Gamma \right) +g^{2}\right] ^{2}}\right\vert ^{2}=\frac{1}{2},
\]%
\[
2\left\vert J_{ba}\right\vert g^{2}\xi \left\vert \sin k_{\mathrm{half}%
}\right\vert =\frac{1}{\sqrt{2}}\left\vert \xi e^{-ik}\left( -2\xi \cos k_{%
\mathrm{half}}+i\Gamma \right) +g^{2}\right\vert ^{2},
\]%
or%
\[
4\left( g^{2}-\xi ^{2}\right) \xi ^{2}\left\vert \sin k_{\mathrm{half}%
}\right\vert ^{2}+2\left( \Gamma -\sqrt{2}\left\vert J_{ba}\right\vert
\right) g^{2}\xi \left\vert \sin k_{\mathrm{half}}\right\vert +4\left( \xi
^{2}-g^{2}\right) \xi ^{2}+\Gamma ^{2}\xi ^{2}+g^{4}=0.
\]%

Under the conditions that $\left\vert J_{ba}\right\vert =2\Gamma$ and $g^{2}=\Gamma \xi$, we have%
\[
2\left( \Gamma -\xi \right) \xi \left\vert \sin k_{\mathrm{half}}\right\vert
^{2}+\left( 1-2\sqrt{2}\right) \Gamma ^{2}\left\vert \sin k_{\mathrm{half}%
}\right\vert +2\left( \xi -\Gamma \right) \xi +\Gamma ^{2}=0.
\]%
Define%
\[
\eta \equiv \frac{\xi }{\Gamma },
\]%
then we have
\[
2\left( 1-\eta \right) \eta \left\vert \sin k_{\mathrm{half}}\right\vert
^{2}+\left( 1-2\sqrt{2}\right) \left\vert \sin k_{\mathrm{half}}\right\vert
+1-2\left( 1-\eta \right) \eta =0,
\]%
or
\[
\left\vert \sin k_{\mathrm{half}}\right\vert =\frac{-\left( 1-2\sqrt{2}%
\right) \pm \sqrt{\left( 1-2\sqrt{2}\right) ^{2}-8\left( 1-\eta \right) \eta %
\left[ 1-2\left( 1-\eta \right) \eta \right] }}{4\left( 1-\eta \right) \eta }.
\]
Define%
\[
\zeta =4\left( 1-\eta \right) \eta,
\]%
then we have
\[
\left\vert \sin k_{\mathrm{half}}\right\vert =\frac{-\left( 1-2\sqrt{2}%
\right) \pm \sqrt{\left( 1-2\sqrt{2}\right) ^{2}-\zeta \left( 2-\zeta
\right) }}{\zeta }.
\]%
The condition for maximum width $\Delta k_{\mathrm{max}}$ is%
\begin{eqnarray*}
\frac{d}{d\eta }\left\vert \sin k_{\mathrm{half}}\right\vert &=&\frac{d \left\vert \sin k_{\mathrm{half}}\right\vert}{%
d\zeta } \frac{d\zeta }{d\eta }
\\
&=&\frac{\left( 2\sqrt{2}-1\right) ^{2}-\zeta -\left( 2\sqrt{2}-1\right)
\sqrt{\left( 2\sqrt{2}-1\right) ^{2}-2\zeta +\zeta ^{2}}}{\zeta ^{2}\sqrt{%
\left( 2\sqrt{2}-1\right) ^{2}-2\zeta +\zeta ^{2}}}\left[ 4\left( 1-2\eta
\right) \right], \\
&=&0,
\end{eqnarray*}%
which is satisfied with
\[
\eta =\frac{1}{2}\Rightarrow \xi =\frac{\Gamma }{2}.
\]%
That is to say, the maximum width $\Delta k_{\mathrm{max}}$ is obtained at $%
\xi =\Gamma /2$ with
\[
\left\vert \sin k_{\mathrm{half}}\right\vert =2\sqrt{2}-1-2\sqrt{2-\sqrt{2}}%
,
\]%
and the maximum FWHM $\Delta k_{\mathrm{max}}$ is%
\begin{equation}
\Delta k_{\mathrm{max}}\equiv \pi -2\sin ^{-1}\left[ 2\sqrt{2}-1-2\sqrt{2-%
\sqrt{2}}\right] \approx 0.81\pi .
\end{equation}

\end{document}